\documentclass[superscriptaddress,secnumarabic,amssymb,nobibnotes]{revtex4-2}

\usepackage{amsmath}
\usepackage{amssymb}
\usepackage{physics}
\usepackage{float}
\usepackage{graphicx}
\usepackage{array}
\usepackage{sublabel}
\usepackage{url}
\usepackage{textcomp}
\usepackage[english]{babel}
\usepackage{multirow}
\usepackage{hyperref}
\usepackage{multirow}
\usepackage{color}

\renewcommand{\vec}[1]{\mbox{\boldmath$#1$}}

\graphicspath{{./}}

\begin{document}

\title{Transverse Confinement of Electron Beams in a 2D Optical Lattice for Compact Coherent X-Ray Sources}%

\author{Arya Fallahi}%
\email[Corresponding author: ]{afallahi@itis.ethz.ch}
\affiliation{Department of Information Technology and Electrical Engineering (D-ITET) Swiss Federal Institute of Technology (ETH Z\"urich), CH-8093 Z\"urich, Switzerland}
\affiliation{IT’IS Foundation, Zeughausstrasse 43, 8004 Zurich, Switzerland}
\author{Niels Kuster}%
\affiliation{Department of Information Technology and Electrical Engineering (D-ITET) Swiss Federal Institute of Technology (ETH Z\"urich), CH-8093 Z\"urich, Switzerland}
\affiliation{IT’IS Foundation, Zeughausstrasse 43, 8004 Zurich, Switzerland}
\author{Lukas Novotny}%
\affiliation{Department of Information Technology and Electrical Engineering (D-ITET) Swiss Federal Institute of Technology (ETH Z\"urich), CH-8093 Z\"urich, Switzerland}
\date{\today}%

\begin{abstract}
	
Compact coherent x-ray sources have been the focus of extensive research efforts over the past decades.
As a result, several novel schemes like optical and nano-undulators for generating x-ray emissions in ``table-top" setups are proposed, developed, and assessed.
Despite the extensive efforts in the past decades, there exists no operational FEL based on optical or electromagnetic undulators.
By combining the particle confinement capability of optical cavities with wiggling motion inside an optical undulator, this paper proposes a new concept for making a compact coherent x-ray source.
The full-wave solution of first-principle equations based on finite-difference time-domain and particle-in-cell (FDTD/PIC) is performed to simulate inverse-Compton scattering (ICS) off both free and confined electrons.
It is shown that the strong space-charge effect in a low-energy electron beam (5\,MeV) is the main obstacle in acquiring coherent gain through the ICS mechanism with a 10-micrometer laser.
Subsequently, it is shown that by confining the electron beam at the field nodes of an optical cavity, the space-charge effect is compensated, and additionally, the ultrahigh charge density enables high FEL-gain at confinement spots.
The full-wave numerical simulations predict enhancement of about three orders of magnitude in the radiation efficiency when ICS is carried out with confined electrons compared to free electrons.
These theoretical results show promising potential as a new scheme for implementing a compact coherent x-ray source.
	
\end{abstract}

\maketitle

\section{Introduction}

Conventional compact x-ray sources include bremsstrahlung tubes \cite{hettinger1958bremsstrahlung}, channeling radiation \cite{ikeda1996parametric}, triboluminescence \cite{camara2008correlation}, and inverse Compton scattering (ICS) sources \cite{esarey1993nonlinear,leemans1996x,rykovanov2014quasi,bech2009hard,graves2014compact}.
The first three make use of non-relativistic electron ($e$-) beams of about the same energy as that of the desired x-ray photon.
Through an interaction between the $e$-beams and a properly designed structure, the energy of non-relativistic electrons is transferred to photons.
While these sources are highly mature and stable technologies, they have certain drawbacks as they: (a) emit with very low directivity, i.e. into $4\pi$ solid angles; (b) emit incoherently, resulting in fluxes that increase only linearly with the number of electrons; (c) are generally not wavelength-tunable; and (d) have difficulties to achieve both high-flux and ultra-small source sizes, as is desirable for various applications.
Currently, the only x-ray sources that do not suffer from these drawbacks are free-electron lasers (FEL) \cite{huang2007review,pellegrini2016physics,saldin1999physics,schmuser2014free}, which, however, require giant facilities.
Thus, research efforts world-wide are focused on the development of more compact x-ray sources.

In an FEL setup, x-ray radiation is produced from interactions between relativistic $e$-beams and an undulator \cite{bonifacio1984collective}.
In the low-gain regime of FEL, incoherent radiation is generated from wiggling motion, whereas, in the high-gain regime, the self-amplification of incoherent radiation due to micro-bunching results in highly coherent and bright radiation.
The majority of the methods currently under investigation adhere to the FEL scheme.
Investigators either try to produce relativistic electrons in a small facility, i.e., perform \emph{compact accelerator} research, or elaborate on realization of short wavelength wiggling motion in a \emph{compact undulator} device.

The ultimate goal in compact accelerator research is to boost the average accelerating gradient.
The use of the plasma environment for particle acceleration engendered the research field of laser-plasma wake-field acceleration (LPWA) \cite{rykovanov2014quasi,geddes2004high,leemans2014multi,leemans2006gev,faure2004laser,malka2002electron}.
This method uses a high-intensity laser that is focused on an atomic gas jet, creating a plasma and in turn inducing extremely high field gradients in the plasma area surrounding the laser. Electrons in the plasma can find the right phase and accelerate to high energies \cite{tajima1979laser,esarey2009physics}.
Gradients of tens of GeV/m in few millimeters of the acceleration path have been demonstrated \cite{leemans2006gev}, and cascades of these interactions have been implemented in a multistage setup \cite{steinke2016multistage}.
Another group of techniques make use of increasing the operating frequency of accelerators.
The higher operation frequency shrinks the device dimensions, which in turn reduces the energy required for particle acceleration, and relaxes limitations on the maximally tolerable fields due to pulsed heating \cite{laurent2011experimental,dal2016experimental,wu2017high}.
Furthermore, higher field emission thresholds, as well as the easier realization of short pulses at high frequencies, assist in obtaining high-gradient accelerators \cite{wang1989rf,loew1988rf}.
Dielectric laser accelerators (DLA) \cite{Peralta2013,Breuer2013,England2014,schonenberger2019generation} as the second method and terahertz (THz) driven linear electron acceleration \cite{Nanni2015,Wong2013,hibberd2020acceleration,xu2020cascaded,othman2020experimental} as the third are promising high-frequency accelerators.
A fourth method, ultrafast accelerator technology, is based on ultrashort pulses employed for electron acceleration \cite{fallahi2016short,zhang2018segmented,huang2016terahertz}.
In addition to the aforementioned methods, there exist numerous other ideas towards high-gradient accelerators, which are investigated in detail.
Some examples are the inverse free-electron laser scheme \cite{curry2018meter}, inverse Cerenkov acceleration \cite{kimura1995laser}, laser plasma acceleration \cite{carbajo2016direct,wong2010direct,wong2017laser}, and beam-driven acceleration \cite{andonian2012dielectric,antipov2012experimental}.

Techniques for the realization of compact undulators include cryogenic undulators \cite{hara2004cryogenic} and optical undulators \cite{bacci2008compact,gallardo1988theory,chang2013high,bacci2006transverse}, where the oscillations in an electromagnetic (EM) wave realize the wiggling motion of electrons.
Sources based on optical undulators are typically referred to as inverse Compton scattering (ICS) sources.
The first ICS x-ray sources were proposed soon after the discovery of lasers in 1963 \cite{arutyunyan1963compton,milburn1963electron} and experimentally demonstrated just one year later \cite{kulikov1964compton}.
The compactness and ability to produce photons of very high energy up to the gamma ($\gamma$-) ray regime is a remarkable peculiarity of ICS sources \cite{phuoc2012all}.
Techniques other than those listed above – e.g., transverse gradient undulators \cite{huang2012compact}, travelling-wave undulator \cite{debus2010traveling}, and interaction with an optical lattice \cite{andriyash2012x} – have also been proposed for x-ray radiation generation.

In addition to the aforementioned schemes, researchers have explored novel approaches for generating coherent x-ray radiation.
Wiggling a modulated electron beam to enable intense superradiant x-ray generation is one example.
The electron modulation can be produced either by field-emitter arrays combined with emittance exchange \cite{graves2012intense}, by field-emission from flat emitter arrays \cite{kartner2016axsis} or by electron beam diffraction from a perfect crystal combined with emittance exchange \cite{nanni2018nanomodulated}.
Moreover, the interaction of an electron beam with a periodic nanostructure leads to radiation through the inverse Smith-Purcell effect \cite{ye2019deep,talebi2017interaction,talebi2016schrodinger}.
Such sources are currently under investigation in various frameworks including metallic gratings realizing surface plasmon polaritons \cite{massuda2018smith}, 2D materials \cite{wong2016towards}, and dielectric nanostructures \cite{roques2019towards}.
Various research institutes around the world are pursuing compact x-ray sources by combining one or more of the above-listed technologies.

All techniques described for the realization of compact x-ray sources are based on the radiation equation in the wiggling motion, namely
\begin{equation}
\lambda_x = \frac{\lambda_u}{2\gamma^2}\left(1+\frac{K^2}{2}\right),
\label{radiationEquationFEL}
\end{equation}
where $\lambda_x$, $\lambda_u$, $K$, and $\gamma$ are the radiation wavelength, undulator period, undulator parameter, and Lorentz factor of the electron beam, respectively.
This equation changes to
\begin{equation}
\lambda_x = \frac{\lambda_0}{4\gamma^2}\left(1+\frac{a_0^2}{2}\right)
\label{radiationEquationICS}
\end{equation}
for ICS interactions with a counter-propagating laser of wavelength $\lambda_0$ and strength parameter $a_0$.
Some techniques elaborate to reach the required Lorentz factor $\gamma$ in a small setup, others to reduce $\lambda_u$ so that the Lorentz factor required for output radiation $\lambda_x$ is decreased.
Nonetheless, accounting for the 3D dynamics of FEL radiation, it can be shown that there are strong fundamental challenges in achieving coherent gain \cite{reiche2010,reiche2006,saldin2010statistical,saldin2008coherence}.
Some examples are meeting the lasing condition $\epsilon < \lambda_x/4\pi$ with $\epsilon$ being the electron beam emittance, preserving the transverse coherence of the beam, delimiting the longitudinal energy spread, and circumventing the space-charge effects.
These effects are safely covered in the theoretical studies only if full-wave solvers based on first-principle equations are used for simulation of the radiation phenomenon.
Therefore, the methodology in this study is using a 3D full-wave solver based on FDTD/PIC to assess the performance of the introduced scheme.

\begin{figure}
	$\begin{array}{c}
	\includegraphics[width=3.0in]{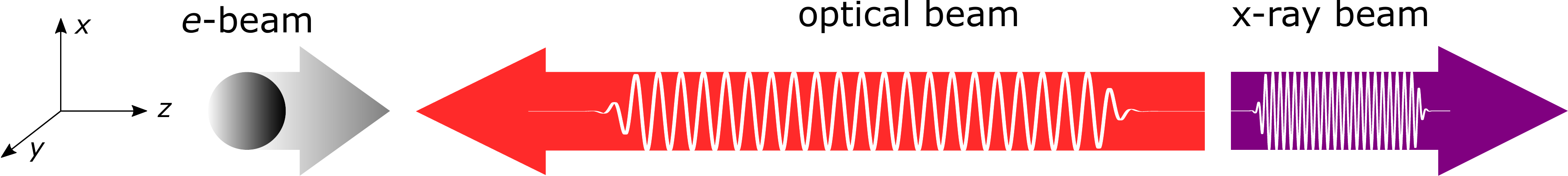} \\
	(a) \\
	\includegraphics[width=3.0in]{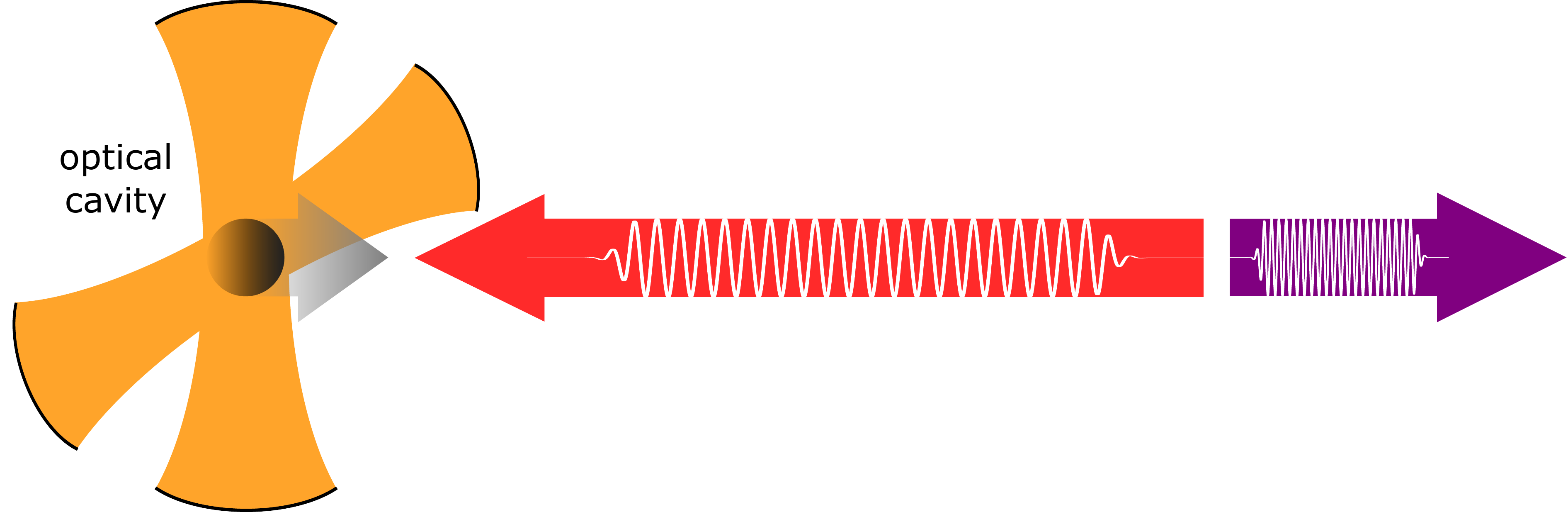} \\
	(b) \\
	\includegraphics[width=3.0in]{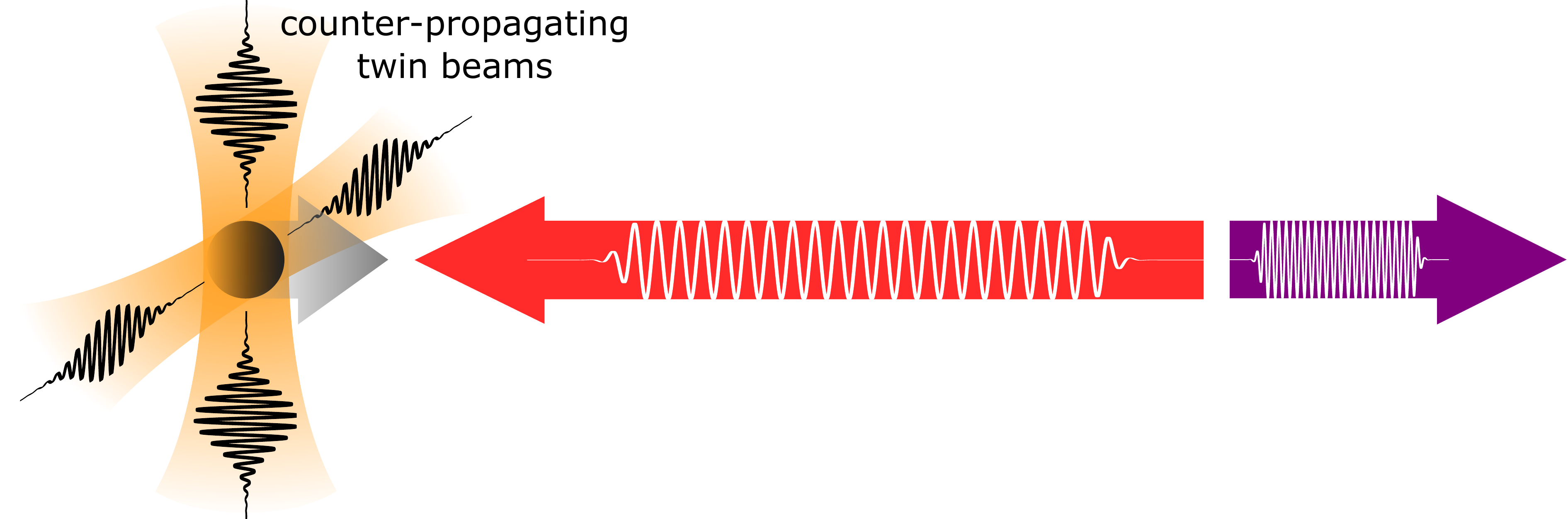} \\
	(c)
	\end{array}$
	\caption{Schematic illustration of different ICS setups: (a) conventional ICS mechanism with optical beam interacting with a free electron beam, and proposed ICS mechanism with optical beam interacting with a confined electron beam inside (b) optical cavity and (c) superposition of counter-propagating beams, respectively.\label{paperConcept}}
\end{figure}

The theory of FEL using an optical undulator has been proposed in the 1980s for the first time \cite{gallardo1988theory,tsunawaki1988optical}.
However, despite the extensive efforts in the past decades, there exists no operational FEL based on optical or EM undulators.
Some studies try to explore the main challenges in free-electron lasing of low-energy electrons using the available simulation tools \cite{bacci2006transverse,chang2013high}.
Fulfilling a detailed investigation on obstacles in FEL operation using optical undulators, similar to the study in \cite{gruner2007design} for LPWA based sources, is out of the scope of this paper.
This paper argues that the strong space-charge forces between electrons are the main impediment in achieving a coherent gain in the radiation.
Afterwards, we aim at establishing a mechanism based on particle beam confinement for relativistic electrons to propose an unconventional approach for tackling the compact coherent x-ray source problem.
The proposed setup is illustrated in Fig.\,\ref{paperConcept}, which shows the conventional optical undulator (Fig.\,\ref{paperConcept}a) setup compared to the new technique.
It is known that charged particles inside a spatially varying field profile are influenced by gradient forces, driving them towards the area of weaker field strengths \cite{gordon1973radiation,ashkin1986observation,chaumet2000time}.
This occurs inside an optical cavity (Fig.\,\ref{paperConcept}b) or when two counter-propagating twin beams impinge on the electron bunch (Fig.\,\ref{paperConcept}c).
It is hypothesized that when the electrons in a beam are transversely confined, many of the existing challenges emanating from transverse motions and repulsive forces are overcome, thereby achieving the coherent-gain regime is alleviated.

\section{Full-wave simulation of FELs}

Numerical simulation of a x-ray FEL in a full-wave fashion based on first-principle equations was for a long time impossible due to the involvement of dramatically multidimensional EM effects.
For instance, the undulator wiggling period is around 1-3\,cm with a total undulator length of 10-100\,m, whereas the radiation wavelength of the radiation source is 0.1-100\,nm.
As a result, very high computation costs are needed to resolve all the involved physical phenomena, which is not practical even with the existing supercomputer technology.

In 2007, the breakthrough theory of simulation in a Lorentz-boosted framework was pioneered by Jean-Luc Vay \cite{vay2007}.
This theory had direct implications on the problem of x-ray FEL simulations, which was further discussed in \cite{fawley2009use}.
The coordinate transformation results in the expansion of bunch size and optical wavelengths and the contraction of the undulator period.
Interestingly, the transformation to electron moving frame transforms these very different length scales to values with the same order of magnitude. 
Consequently, the length of the computation domain is reduced to slightly larger than the bunch dimensions making the full-wave simulation numerically feasible.
Based on this principle, the code MITHRA was developed which provides a full-wave simulation tool for FEL mechanism \cite{fallahi2016mithra}.
This code is widely benchmarked against standard simulation tools in FEL and accelerator physics \cite{fallahi2020MITHRA2A} and is currently incorporated into the OPAL platform \cite{alba2020opal}.
All simulations presented in this paper are performed using MITHRA.
The data for reproducing the presented results in this paper are open-source and available under \href{https://github.com/aryafallahi/mithra}{https://github.com/aryafallahi/mithra}.

Typical finite-difference time-domain particle-in-cell (FDTD/PIC) codes like WARP and OSIRIS employ the Yee algorithm to solve Maxwell equations and a symplectic algorithm like Boris scheme to solve for the particle motion.
Such a framework is not able to capture the Coulomb forces between the charged particles, i.e. the so-called space-charge effect.
To this end, the Poisson equation needs to be solved in conjunction with the Maxwell and motion equations.
MITHRA implements the FDTD/PIC algorithm on potential equations, namely
\begin{align}
\label{WaveA}
\nabla^2\vec{A} - \frac{1}{c^2} \frac{\displaystyle \partial^2 \vec{A}}{\displaystyle \partial t^2} & = -\mu_{0} \vec{J} \\
\label{WaveF}
{\nabla}^{2}\varphi- \frac{1}{c^2} \frac{\displaystyle \partial^2 \varphi}{\displaystyle \partial t^2} & = -\frac{\displaystyle \rho}{\displaystyle \varepsilon_{0}},
\end{align}
where $\vec{A}$, $\phi$, $\vec{J}$ and $\rho$ are the time- and space-dependent magnetic vector potential, electric scalar potential, current density, and charge density, respectively.
The electric and magnetic fields are extracted from potential fields to update the particles' phase-space coordinates.
Subsequently, the charge and current densities, $\vec{J}$ and $\rho$, are updated according to the motion in the latest time step.
The updated values for $\vec{J}$ and $\rho$ are inserted in (\ref{WaveA}) and (\ref{WaveF}) to solve for the backaction of particles' motion on the EM fields.
Through such an implementation, the space-charge effect is covered in the simulations (Example 4 in \cite{fallahi2020MITHRA2A}).
Although the space-charge effect is negligible in conventional x-ray FELs owing to the ultra-relativistic beam energies, as will be seen later, it dramatically influences the operation of compact x-ray sources.

In addition to the space-charge effect, the EM recoil and quantum effects also should be addressed.
In \cite{jianyuan2018structure}, it is shown that the finite-difference formulation for potential equations coupled to a symplectic algorithm for the equation of motion preserves the symplectic two-form associated with the original Hamiltonian systems, similar to an exact solution.
In other words, such an algorithm should also capture the EM recoil effect.
On the other hand, the classical treatment of electron-wave interaction within MITHRA does not allow for any consideration of quantum mechanical effects.
The ratio $\rho_1 = \hbar\omega/\gamma mc^2$, with $\omega = 2\pi c/\lambda_x$, representing the amount of quantum recoil due to each photon emission, quantifies the quantum mechanical effects.
Furthermore, $\rho_2 = (\hbar\omega/2 \rho_{FEL} \gamma mc^2)^2$ with $\rho_{FEL}$ being the FEL parameter, estimates the level of recoil influence on the gain process \cite{bonifacio2005quantum,bonifacio2006quantum}.
The simulated cases in this paper hold the condition $\rho_1 \ll 1$ and $\rho_2 \ll 1$ thus the use of MITHRA software and its implemented formulation is justified.

\section{Challenges in lasing of optical undulators}
 
The Pierce or FEL parameter, $\rho_{FEL}$ describes the conversion efficiency from the electron beam power
into the FEL radiation power.
In a one-dimensional ideal case, where detrimental effects originating from energy spread, emittance, diffraction, space-charge, and time-dependence of the beam and radiation profile are neglected, the FEL parameter for planar undulator can be derived from \cite{schmuser2014free}
\begin{equation}
\rho_{FEL} = \frac{1}{4\gamma}\left(\frac{\mu_0 e^2}{2 \pi^2 m_e} \hat{K}^2 \lambda_u^2 n_e \right)^{1/3},
\label{pierceParameterA}
\end{equation}
where $\hat{K} = K ( J_0(\xi) - J_1(\xi) )$ with $J_0$ and $J_1$ being first and second order Bessel functions and $\xi = K^2/(4+2K^2)$.
In addition, $n_e$ stands for the electron density in the beam.
For an optical undulator, after considering the transformations $\lambda_u \leftrightarrow \lambda_0/2$ and $K \leftrightarrow a_0$, the equivalent FEL parameter is obtained from 
\begin{equation}
\rho_{FEL} = \frac{1}{4\gamma}\left(\frac{\mu_0 e^2}{8 \pi^2 m_e} \hat{a}_0^2 \lambda_0^2 n_e \right)^{1/3}.
\label{pierceParameterB}
\end{equation}
The main goal in utilizing optical undulators is to achieve a targeted $\lambda_x$ using a smaller $\lambda_u$ and $\gamma$.
By accounting for a constant $\lambda_x$, equations (\ref{pierceParameterB}) and (\ref{radiationEquationICS}) yield the following equation for the FEL parameter:
\begin{equation}
\rho_{FEL} = \left(\frac{\mu_0 e^2}{32 \pi^2 m_e} \frac{\hat{a}_0^2 \lambda_x^2}{(1+a_0^2/2)^2} \gamma n_e \right)^{1/3}.
\label{pierceParameterC}
\end{equation}
The laser strength parameter $a_0$ is often constant and set close to one to procure maximum efficiency on one hand and avoid disadvantageous nonlinear effects on the other.
Hence, it can be deduced that obtaining the same coherent radiation efficiency ($\rho_{FEL}$) for low-energy electrons requires a proportionally higher electron density ($n_e$).
This requirement is completely contrary to the space-charge effect, which becomes considerably stronger in the low-energy regimes.
As a result, an electron density which may be easily achievable for electron beams with GeV energy can be difficult to realize at MeV electron energies let alone the thousand times higher density dictated by equation (\ref{pierceParameterC}).
In addition to the realization of the required electron density, maintaining this high density throughout the whole interaction length is of utmost importance.

To demonstrate this hypothesis, an exemplary FEL interaction with parameters tabulated in table \ref{ICSparameters} is taken into account.
\begin{table}
	\caption{FEL parameters configuration with optical undulator.}
	\centering
	\begin{tabular}{|c||c|}
		\hline
		FEL parameter & Value \\ \hline \hline
		Current profile & Uniform \\ \hline
		Bunch RMS size ($\sigma_x = \sigma_y$) & 5\,{\textmu}m \\ \hline
		Bunch length ($\sigma_z$) & 10\,{\textmu}m \\ \hline
		Bunch charge ($Q$) & 0.32\,pC \\ \hline
		Bunch energy ($E$) & 5.12\,MeV \\	\hline
		Bunch current ($I$) & 19.2\,A \\ \hline
		Longitudinal momentum spread ($\sigma_{\gamma \beta_z}$) & $2.5\times 10^{-3}$  \\ \hline
		Normalized emittance ($\epsilon$) & 50 nm-rad \\	\hline
		Laser wavelength ($\lambda_0$) & 10\,{\textmu}m \\ \hline
		Laser strength parameter ($a_0$) & 1.0 \\ \hline
		Pulse duration ($cT$) & 18\,mm \\ \hline
		Laser pulse type & flat-top \\ \hline
		Radiation wavelength ($\lambda_x$) & 37.6\,nm \\ \hline
	\end{tabular}
\label{ICSparameters}
\end{table}
The considered values are chosen based on the state-of-the-art electron gun as well as CO$_2$ laser technologies.
The interaction is simulated using the code MITHRA with and without considering the space-charge effect.
The analysis without space-charge effect is performed by setting $\phi = \rho = 0$ in equation (\ref{WaveF}).
The electron and optical beam propagation direction is assumed to reside along the $z$-axis and the optical beam is linearly polarized along the $y$-axis.
In the simulations, the electron bunch is modeled by $2\times10^6$ particles each representing one electron, and a shot noise is added over the particle distribution.
The co-moving computational domain around the bunch is discretized into 45 million pixels, which leads to around 16 hour simulation time on 128 parallel nodes.

\begin{figure}
	$\begin{array}{cc}
	\includegraphics[width=3.2in]{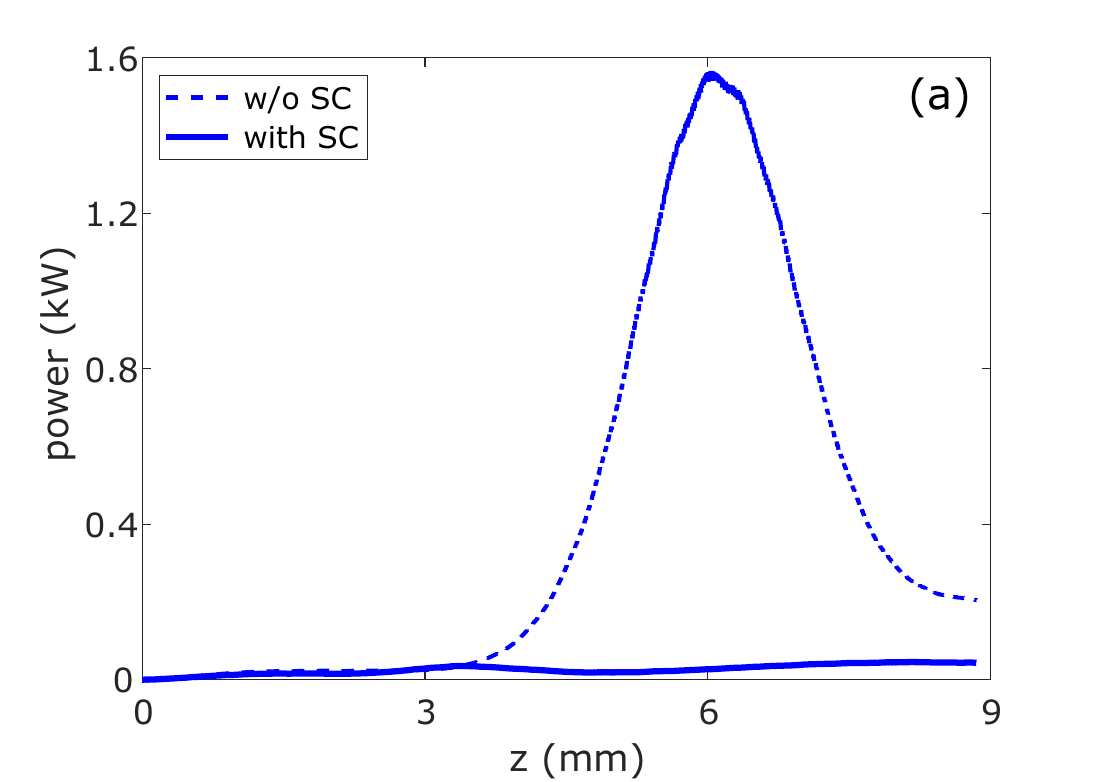} &
	\includegraphics[width=3.2in]{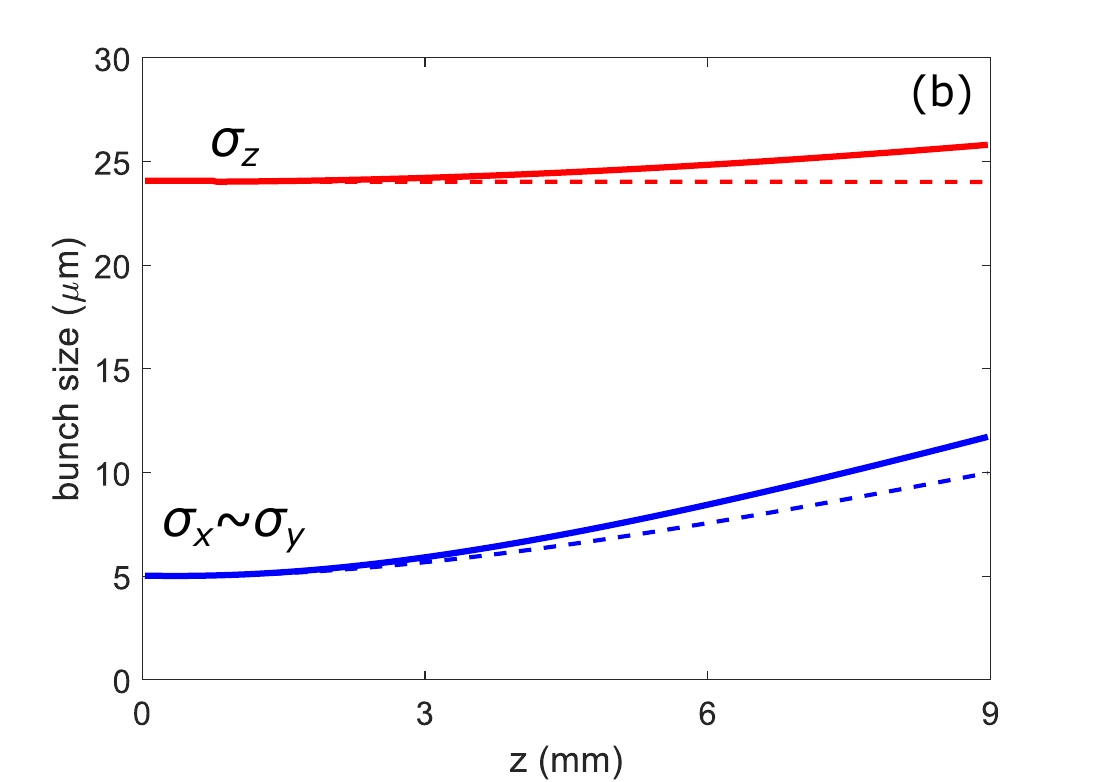}
	\end{array}$
	\caption{(a) Radiated power and (b) bunch dimensions, namely length (red) and transverse size (blue), throughout the ICS interaction summarized in table \ref{ICSparameters}. The results with (solid lines) and without space-charge (dotted lines) are shown. \label{ICSResult}}
\end{figure}

Fig.\,\ref{ICSResult}a shows the comparison of the radiated power from the bunch for the two simulated cases.
The predicted radiation efficiency for the simulation without space-charge consideration is $3.0\times10^{-5}$, which reduces to $9.1\times10^{-7}$ when the effect is considered.
It is observed that the minute radiation gain obtained in the analysis without space-charge is completely suppressed when the Coulomb repulsion of electrons is taken into account.
Comparing the bunch sizes and lengths in the two analyses, which is shown in Fig.\,\ref{ICSResult}b, confirms that the space-charge forces intensify the bunch expansion.
As indicated in Fig.\,\ref{ICSResult}b, the longitudinal size of the beam increases by 7\% due to the space-charge effect.
Such a large variation in the bunch length strongly degrades the self-modulation of the bunch at the correct wavelength.
Hence, the coherent gain is hampered, and thereby the FEL radiation efficiency is dramatically reduced.
Our results indicate that a mechanism is required that substantially reduces the gain-length such that lasing of the low-energy electrons is enabled before the bunch longitudinal expansion degrades the amplification process.
To this end, higher charge densities need to be realized and additionally preserved throughout the interaction.
This is in line with the hypothesis on scaling FEL operation to small undulator periods discussed in the previous section.

\section{Confinement of electron beams traversing optical cavities}

It is known that when a particle is exposed to spatially varying EM fields, a so-called \emph{gradient force} or \emph{ponderomotive force} influences the particle motion \cite{landau2013electrodynamics,gaponov1958potential}.
This pondermotive force is derived from
\begin{equation}
\vec{F}_p = - \frac{e^2}{4m_e\omega^2}\nabla (E^2),
\label{ponderomotiveForce}
\end{equation}
where $E$ denotes the time-varying, i.e. oscillating electric field and $\omega$ is the oscillation frequency.
This equation suggests that a charged particle in an inhomogeneous oscillating field not only oscillates at the $\omega$ frequency but is also accelerated by $\vec{F}_p$ toward the weak field locations.
When an ensemble of charged particles resides in the standing wave of an optical or EM cavity, owing to the existence of points with minimum and maximum field strengths, ponderomotive forces reshape the bunch distribution.
More accurately, these forces push the particles to field nodes and pull them away from the field antinodes.
Similarly, it is expected that when an electron beam traverses an optical cavity, the transverse distribution of the electrons is influenced by the gradient forces in an analogous fashion.

\begin{figure}
	$\begin{array}{cc}
	\includegraphics[width=3.2in]{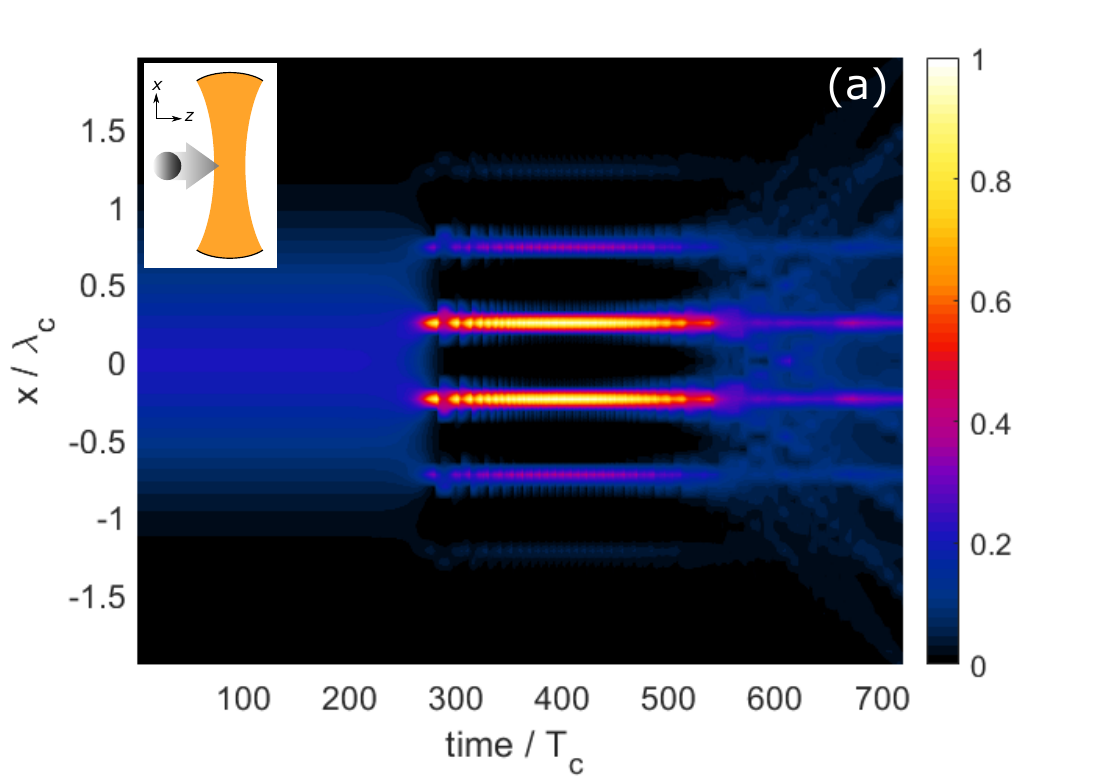} &
	\includegraphics[width=3.2in]{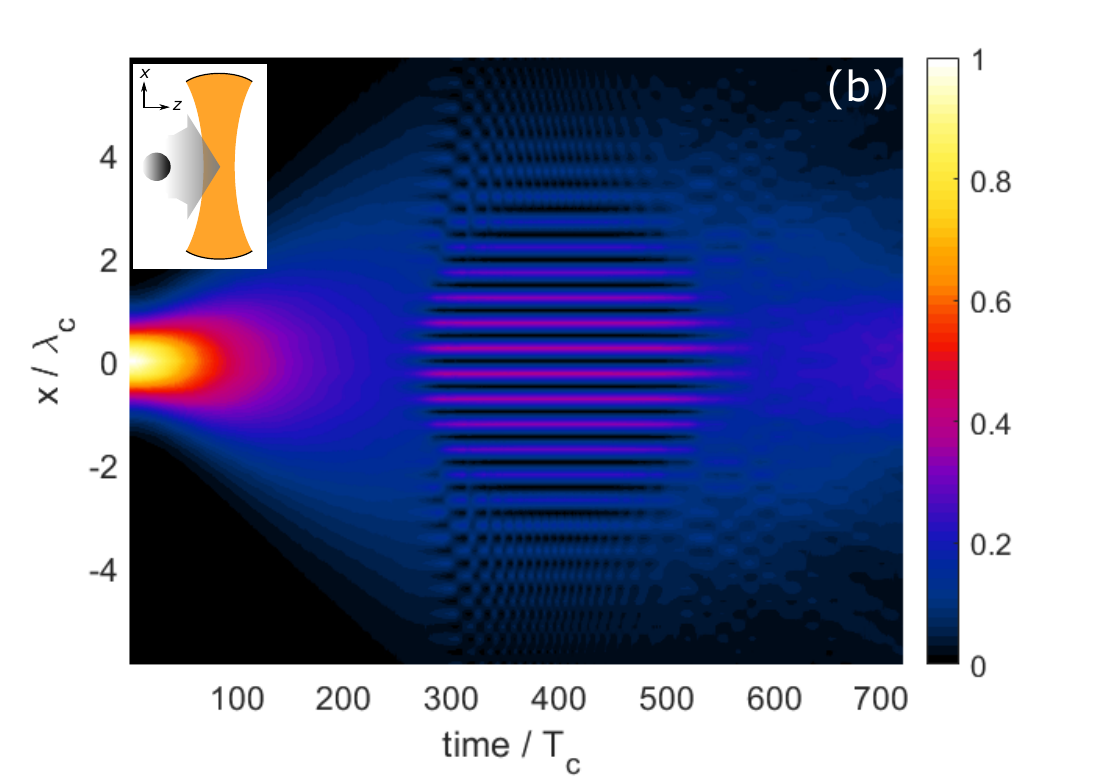}
	\end{array}$
	\caption{ Evolution of normalized charge density when (a) an ideal beam ($\epsilon =0$) and (b) a diverging beam enters an optical cavity. The charge density is normalized to the maximum density in the considered space and time window. \label{beamCavityIllustration}}
\end{figure}

Fig.\,\ref{beamCavityIllustration}a illustrates the variations in normalized charge density of an ideal beam (zero emittance) when it passes through a cavity.
This visualized data is obtained through particle-in-cell simulation of electrons experiencing cavity fields which are formulated as a standing-wave Gaussian field distribution (see Appendix).
Without loss of generality, a 2D interaction problem, as schematically shown in the figure subset, is considered here.
In this 2D case, confinement happens only along one transverse direction ($x-$axis in Fig.\,\ref{beamCavityIllustration}).
A second simulation tackles the extreme case, in which an exploding beam with high emittance enters the cavity.
The results of this simulation are illustrated in Fig.\,\ref{beamCavityIllustration}b.
Note that these 2D simulations are presented to qualitatively examine the bunch evolution inside the optical cavity yet, for the sake of brevity, the detailed parameters for the beam and cavity are not outlined.
In both ideal and extreme cases the confinement of particles in specific regions coinciding with field nodes is well observed.

Originated from this confinement effect, three closely interrelated phenomena happen simultaneously, which are advantageous for achieving the coherent-gain.
First, the confinement effect prevents the increase in beam transverse size due to the transverse energy spread, thereby minimizing the beam effective emittance.
It should be emphasized that the electrons are confined inside a potential well produced by the ponderomotive forces.
Athough the intirinsic emmitance of the beam is conserved, since the microbeams are no more free beams the beam divergence is not directly related to the beam emittance.
Considering the effective emittance as the product of the beam divergence with its transverse size, this parameter is reduced  inside the cavity fields.
This dramatic reduction in the effective emittance facilitates meeting the lasing-condition, $\epsilon < \lambda_x/4\pi$, and consequently makes the electron beam a perfect candidate for an FEL mechanism.

Second, if confinement is enforced along both transverse directions, the ponderomotive force suppresses the transverse expansion of the beam size due to Coulomb forces.
Although this expansion marginally reduces the charge density (see Fig.\,\ref{ICSResult}b), its prevention is important when high charge densities are realized.

Third, and most importantly, is the realization of beams with considerably higher charge densities.
As shown in Fig.\,\ref{beamCavityIllustration}a and \ref{beamCavityIllustration}b, the ponderomotive force divides a single low-charge density beam at the cavity entrance to several beams with intensified charge densities.
Inspired by the name \emph{microbunches} in FEL physics, throughout this paper, these beams inside the cavity are named as \emph{microbeams}.
As postulated in the previous sections, the higher charge density of each microbeam augments their aptitude for coherent-gain in the radiation.
As a conclusion, a freely propagating electron beam passing through the cavity transforms into several confined microbeams which are more suitable candidates for coherent radiation compared to the original beam. 

\section{Inverse Compton Scattering off a confined electron beam}

\begin{figure}
	$\begin{array}{cc}
	\includegraphics[width=3.2in]{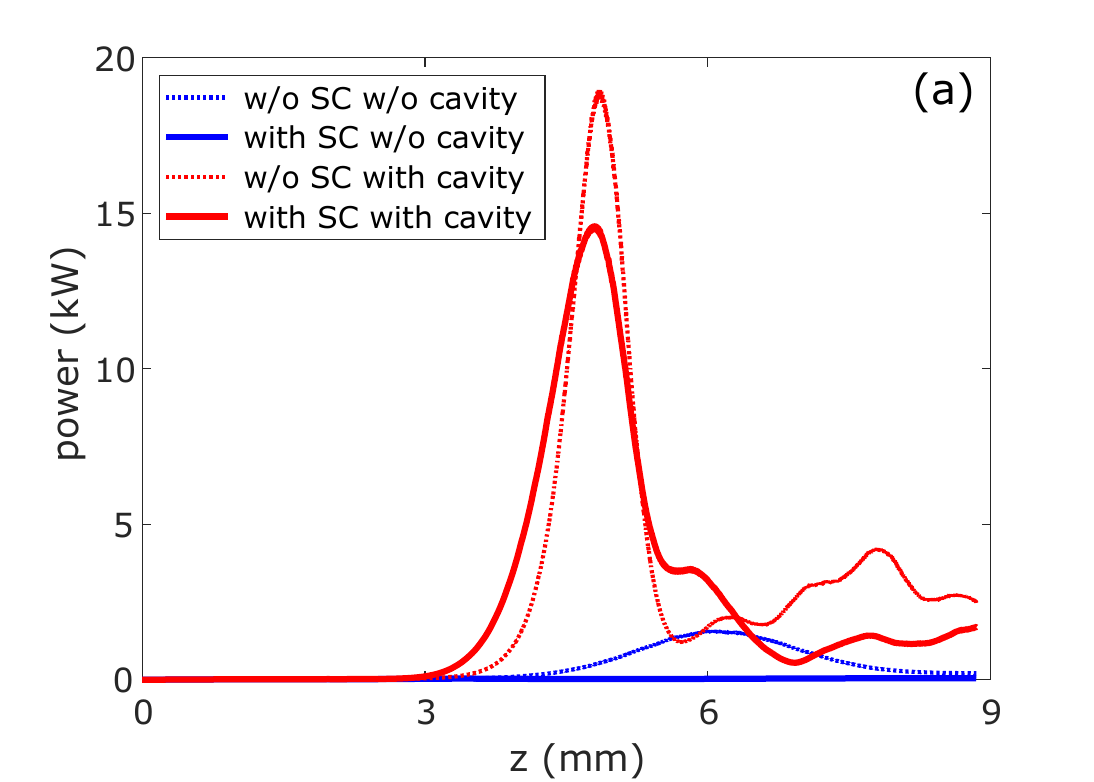} &
	\includegraphics[width=3.2in]{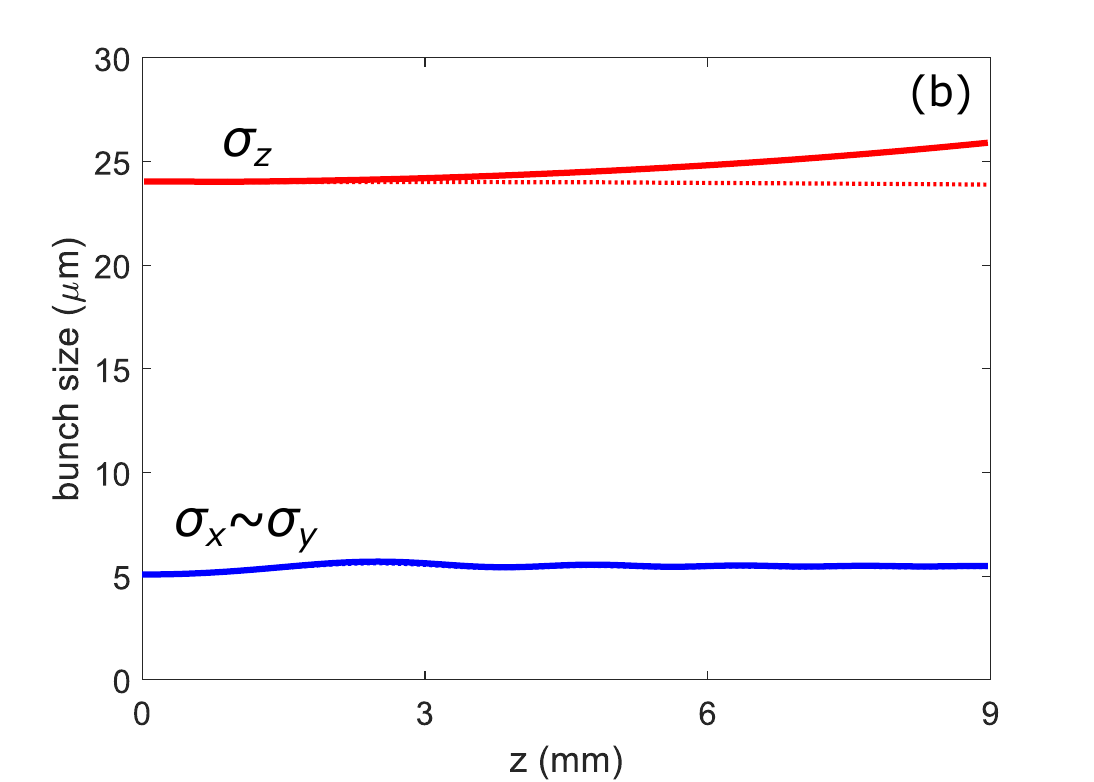} 
	\end{array}$
	\caption{(a) Radiated power and (b) bunch dimensions resulting from the ICS interaction with confined electron beams whose specifications are tabulated in table \ref{ICSparameters} and \ref{cavityParameters}. The results with (solid lines) and without space-charge (dotted lines) are compared.} \label{trappedICSResult}
\end{figure}

To verify this hypothesis, the same ICS simulation as in table\,\ref{ICSparameters} is performed but with the fields of standing-waves superposed on the field of the counter-propagating ICS laser.
Here, two cavities configured along $x$- and $y$-axes are taken into account such that the particle confinement happens in both transverse directions.
The cavity beam parameters are listed in table\,\ref{cavityParameters}.
The cavity wavelength ($\lambda_c = 11.8$\,{\textmu}m) and the center position with respect to the starting point of bunch undulation ($c\delta t = 8.4$\,mm) are obtained through optimization sweeps.
A physical interpretation of these values will be explained in the next sections.
Moreover, the two cavities are linearly polarized with a magnetic field along the beam direction, i.e. $B_z$ polarized.
This is a critical requirement for the cavity configuration because an electric field along the beam traveling direction disturbs the longitudinal motion of the electrons and consequently the formation of microbunches.
The total radiated power at $\lambda_x = 37.6$\,{\textmu}m as well as the bunch dimensions for simulations with and without space-charge are shown in Fig.\,\ref{trappedICSResult}.
The obtained results for final radiation are compared against the simulation without cavity fields, namely conventional ICS off free electrons (Fig.\,\ref{ICSResult}a).

\begin{table}
	\caption{Parameters of the cavity beams.}
	\centering
	\begin{tabular}{|c||c|}
		\hline
		Cavity parameter & Value \\ \hline \hline
		Wavelength ($\lambda_c$) & 11.8\,{\textmu}m \\ \hline
		Beam size along electron beam ($w_{0z}$) & 7.4\,mm \\ \hline
		Beam size normal to electron & \multirow{2}{*}{0.1\,mm} \\
		beam ($w_{0x}$ or $w_{0y}$) & \\ \hline
		Beam strength parameter ($a_{0c}$) & 0.07 \\ \hline
		Center position ($c\delta t$) & 8.4\,mm \\	\hline
	\end{tabular}
	\label{cavityParameters}
\end{table} 

Several effects are observed in the results which are in agreement with the previously addressed predictions.
First, the transverse expansion of the beam due to beam intrinsic emittance and space-charge effects are totally suppressed as seen in Fig.\,\ref{trappedICSResult}b.
Second, the results without space-charge consideration show that the intensified charge density leads to smaller gain-length and ultimately higher FEL efficiency.
Third, there is a considerably less detrimental effect due to space-charge fields.
Due to the fast coherent gain process, the FEL radiation is saturated before space-charge forces expand the longitudinal size of the beam.
As a result, the ultimate radiation efficiency of the ICS source with confined electrons increases to $2.8\times10^{-4}$, which is close to conventional FEL sources.
Comparing the source efficiencies with and without cavity fields reveals an enhancement factor of $\approx 307$ owing to the confinement effect of electrons.

\begin{figure*}
	\includegraphics[width=6.8in]{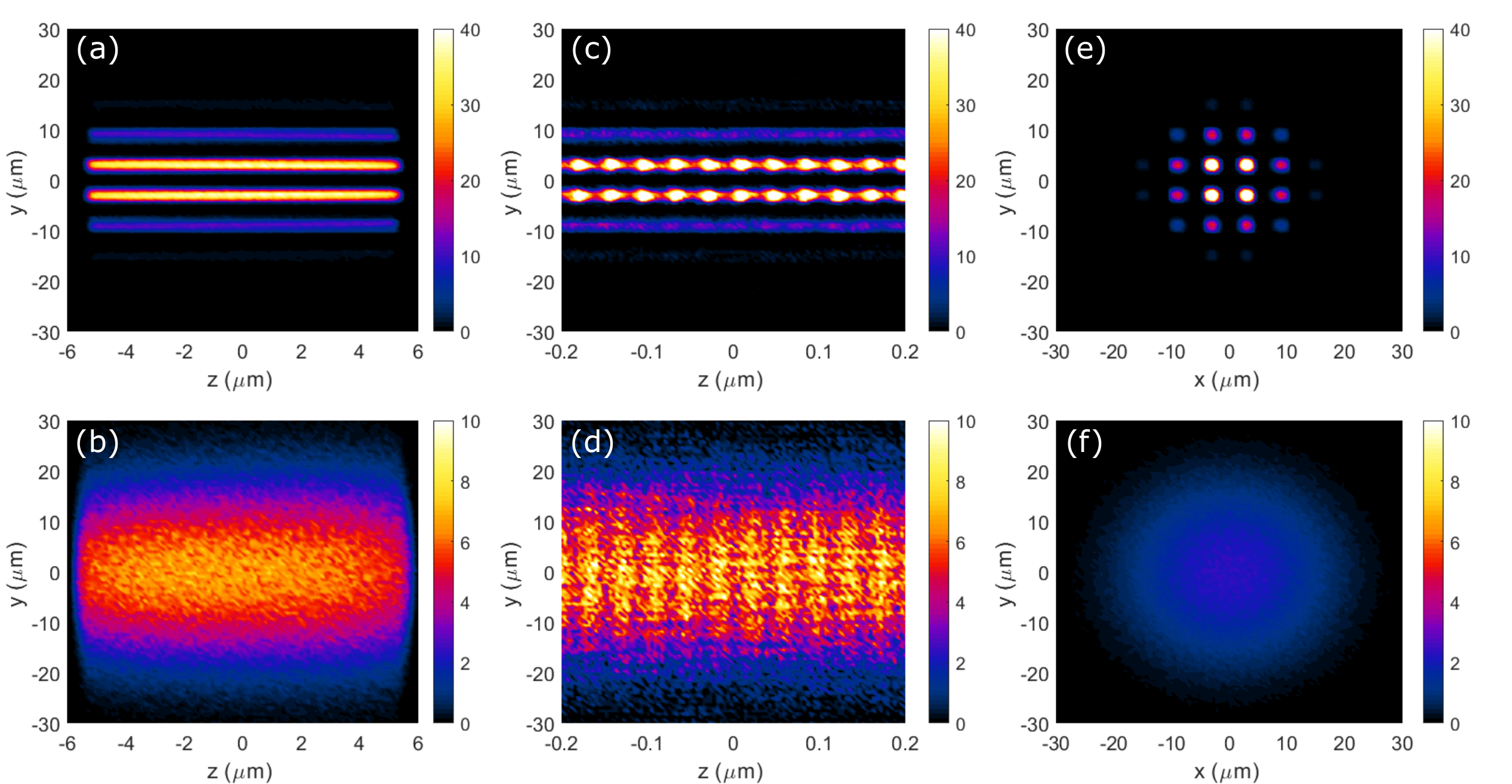}
	\caption{Bunch profiles at the maximum radiation instant: In (a-d) $dN/dydz$ [$10^3$/{\textmu}m$^2$] with $N$ the number of electrons is plotted. (c) and (d) are the zoomed-in profiles of (a) and (b) respectively, to show the microbunching. In (e) and (f) $dN/dxdy$ [$10^3$/{\textmu}m$^2$] is plotted. (a), (c) and (e) are profile visualizations for simulations with cavity fields and (b), (d) and (f) visualize the bunch profile for simulations without cavity fields. \label{bunchProfile}}
\end{figure*}

Scrutiny of the bunch and radiation profile sheds more light on the radiation mechanism of a confined electron bunch.
Figure\,\ref{bunchProfile} shows the bunch profile at the maximum radiation instant for simulations with and without cavity fields.
For the visualizations here, results of the simulations capturing the space-charge effect are considered.
Note that the bunch profiles viewed along the $x$- and $y$-axis are analogous owing to the inherent symmetry of the problem.
In correspondence with the described hypothesis, the initial beam is divided into several microbeams (Fig.\,\ref{bunchProfile}e and f), that enable shorter gain-length due to higher charge density.
This results in overcoming space-charge forces and consequently the rapid formation of microbunches in each microbeam (Fig.\,\ref{bunchProfile}a-d).
The radiations in the microbunches then add up coherently and lead to the observed enhancement in the total radiation.

\begin{figure*}
	\includegraphics[width=6.8in]{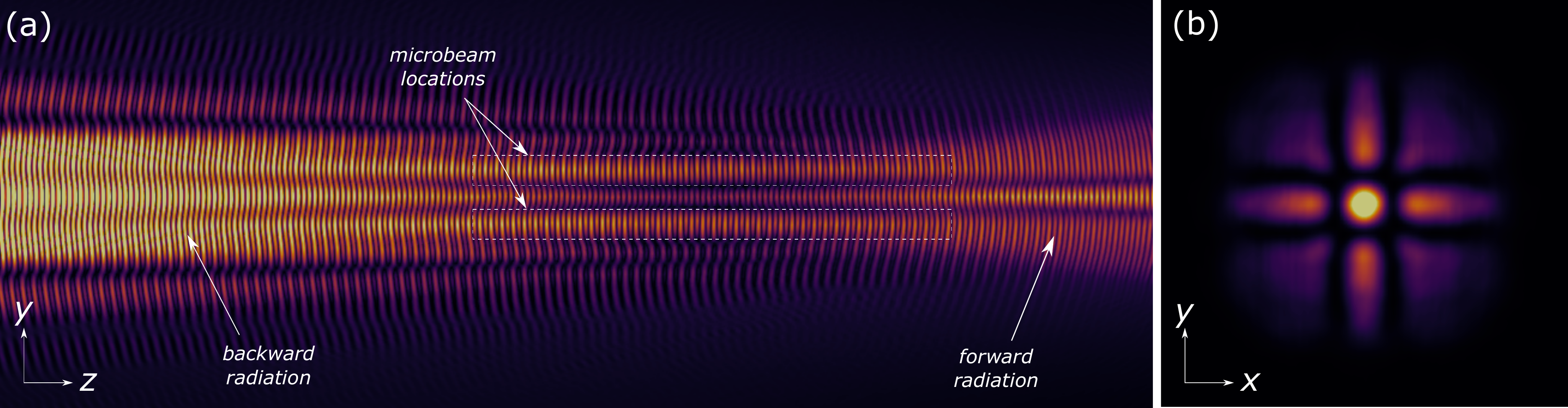}
	\caption{Radiation profiles in the bunch co-moving frame at the maximum radiation instant for simulations with cavity fields and space-charge effect: (a) shows the magnitude of the electric-field viewed from $x$-axis at the cavity nodes and (b) illustrates the radiated power at a detector plane located in front of the bunch \label{radiationProfile}}
\end{figure*}

In Fig.\,\ref{radiationProfile}a, the radiation profile in the ICS from the confined electron beam in its co-moving coordinate frame is visualized.
The radiated field is depicted on the cavity nodes as well as on a plane residing 5.7\,{\textmu}m in front of the bunch center.
The results show that the forward-propagating radiation becomes as strong as the coherent backward propagating radiation at the maximum radiation instant.
The output beam from the proposed source has a different beam profile from conventional FEL sources.
Since multiple radiation sources are produced, radiation main-lobe and side-lobes emerge as in phased array antennas.
This is observed in visualized power profile in Fig.\,\ref{radiationProfile}b.

The obtained results numerically confirm the promising potential for ICS with confined electron beam as a new generation for x-ray sources.
In the next sections, the anticipated characteristics of the proposed x-ray source are discussed.

\section{Coherence properties of the radiation}

\begin{figure}
	$\begin{array}{cc}
	\includegraphics[width=3.2in]{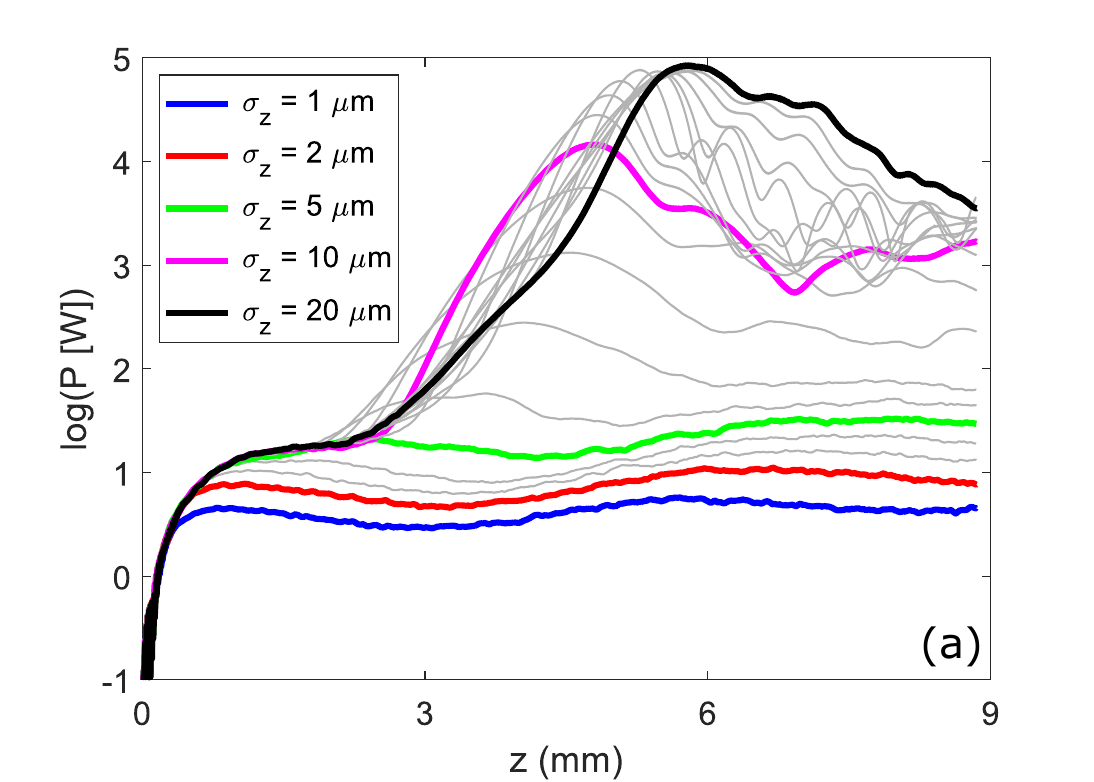} &
	\includegraphics[width=3.2in]{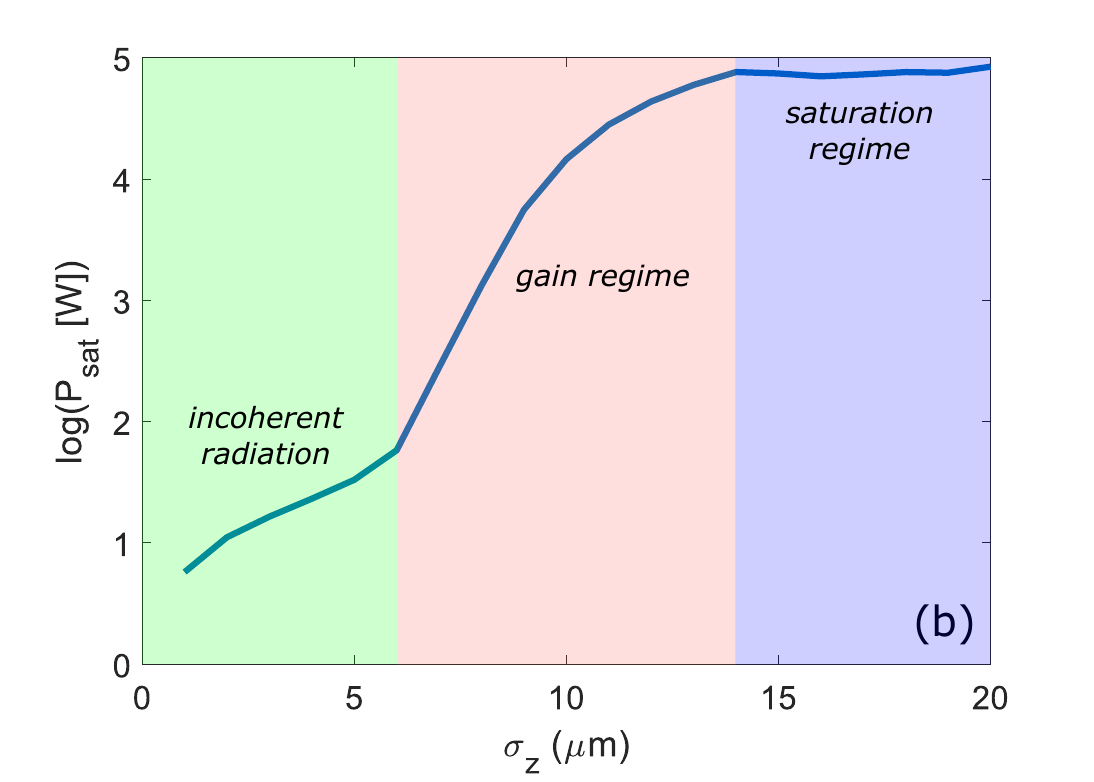} \\
	\end{array}$
	\caption{(a) Radiated power versus interaction length for various bunch lengths and (b) saturation (maximum) radiation power versus bunch length for the ICS interaction from confined electrons in a cavity. \label{radiationvsBunchLength}}
\end{figure}

The presented visualization of the bunch profile illustrated qualitatively the occurrence of microbunching in each microbeam.
Notwithstanding, such a data visualization does not demonstrate the event of coherent gain in the output radiation.
To demonstrate the coherent gain in radiation, and thereby theoretically proving the functioning of a compact \emph{coherent} x-ray source, the maximum radiated power was investigated when the bunch length is varied.
By changing proportionally the total number of electrons, a constant average charge density is maintained throughout the sweep.
The bunch length is directly proportional to the number of microbunches.
Thus, the variation of the saturation (i.e. the maximum) power in terms of the bunch length shows the degree of coherent gain occurring during the interaction.
Fig.\,\ref{radiationvsBunchLength} shows the result of this study on a logarithmic scale.
It can be seen that in the small bunch length regime, the saturation power increases slightly with the bunch length.
The increase rate becomes drastically stronger for longer bunch lengths after the coherent gain is triggered.
Eventually, the radiated power is saturated and stays constant with the bunch length.
This observation is in agreement with the free-electron laser principle and confirms the coherent buildup of the radiation from each microbunch.

It is known from FEL physics that the transverse and longitudinal coherence are two strongly interdependent properties \cite{saldin2010statistical,saldin2008coherence}.
In a conventional FEL setup, the radiation of individual microbunches is coherently amplified if the transverse coherence of the radiation is preserved throughout the beam cross-section.
This raises an intricate question on the coherent radiation from microbunched microbeams in the proposed scheme: why does the microbunching in all microbeams take place coherently in-phase (see. Fig.\,\ref{bunchProfile}c)?

\begin{figure}
	\includegraphics[width=3.2in]{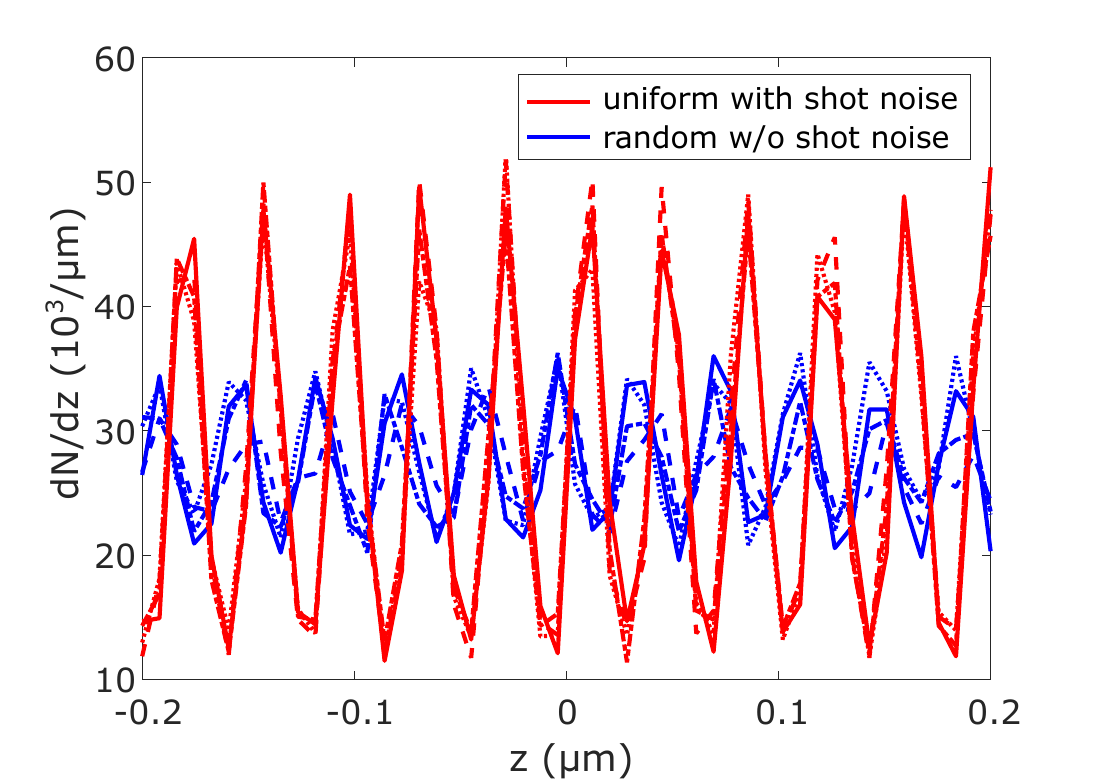}
	\caption{ $dN/dz$\,[10$^3$/{\textmu}m] as the longitudinal charge density is depicted along the bunch. The four red and blue lines show the microbunching of the four central microbeams when the particle distribution is initialized according to shot noise implementation algorithm and random generation, respectively. \label{randomBunchGeneration}}
\end{figure}

Note that in the presented simulations, shot noise is added over uniformly distributed electrons as the trigger source for coherent radiation \cite{penman1992simulation,fawley2002algorithm}.
This shot noise is a $z$-dependent function, which may raise the suspicion that the observed transverse coherence among all the microbeams is the consequence of this assumption.
To test this effect, a simulation is performed in which the electrons are distributed randomly according to the random number generator of the simulation platform, with no additional shot-noise.
It is known that such a bunch generation algorithm leads to unrealistic clustering of the electrons and consequently deviations from the correct simulation results \cite{pellegrini2016physics}.
However, its effect on the microbunch formation assists in clearing up the aforementioned ambiguity.
The result is shown in Fig.\,\ref{randomBunchGeneration} where the particle density along the traveling direction is depicted for the four central microbeams (see Fig.\,\ref{bunchProfile}e).
As observed in the simulations, despite the random generation of particles, the microbeams are all microbunched coherently in-phase, resulting in transverse coherent radiation.

There are two primary reasons leading to the observed effect.
First, we know from FEL physics that the microbunch phase is mainly determined during the initial gain lengths (\cite{schmuser2014free} pages 79-81).
After the optimization process, it was always observed that the position of the cavity with respect to the ICS interaction point is set such that wiggling motion is started before the formation of microbeams.
As a consequence, during the time when the microbunch phase is determined no transverse modulation exists, which results in a constant microbunching phase throughout the whole beam.
This phasing does not change after the formation of microbeams and as a result, all microbeams are micorbunched coherently in phase.
The second reason is linked to the coupling between the microbeams through the radiated fields.
As observed in the radiated field profile (Fig.\,\ref{radiationProfile}a), the radiation from each microbeam strongly diffracts because of the small beam size of the microbeams.
This beam diffraction effect causes a microbeam to be influenced by the superposition of radiated fields from the whole bunch.
Therefore, the microbunching of individual microbeams is a collective effect in the whole bunch similar to a conventional FEL mechanism.
As a consequence, transversely coherent microbunching is anticipated in the proposed interaction.

\begin{figure}
	\includegraphics[width=3.2in]{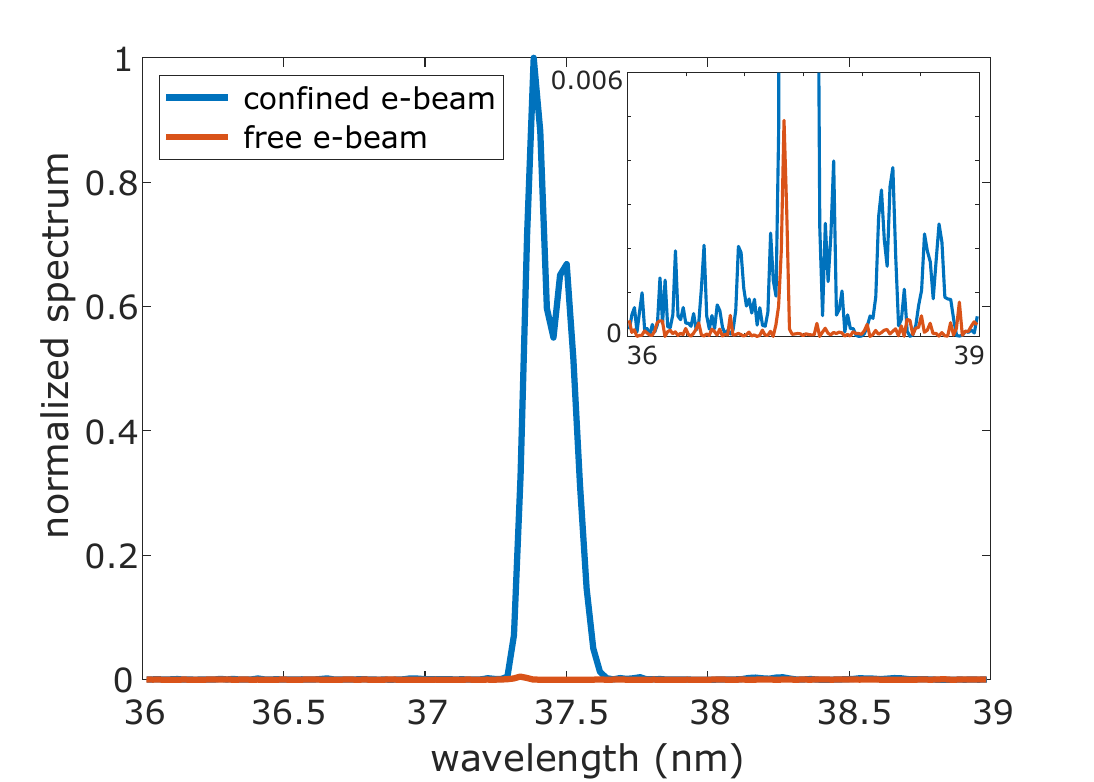}
	\caption{Radiation spectra for the on-axis radiations for ICS off confined and free electron beams. Both spectra are normalized to the maximum value of the radiation spectrum in confined $e$-beam case. \label{spectrum}}
\end{figure}

Coherence time is an important property of any radiation source including the proposed x-ray source.
Here, the coherence time ($\tau$) is approximated as the inverse of spectral bandwidth for the on-axis radiation.
In Fig.\,\ref{spectrum}, the radiation spectra of on-axis radiations are depicted for ICS off free electrons (no cavity fields) and confined electrons.
It is seen that the spectrum of the ICS source is slightly broadened when electrons are confined in the cavity fields.
Cavity fields induce a transverse oscillating motion to the electron beams around the cavity nodes (see Fig.\,\ref{beamCavityIllustration}a).
This means that the particles are not transversely stationary at the cavity nodes but are oscillating around the nodes.
As a result of the two simultaneously happening oscillations due to ICS and cavity beams and the non-linear effects emanated from the longitudinal motion of electrons, a wave-mixing process takes place which leads to a broadening of the radiation spectrum.
In other words, there exists a trade-off between the output power and spectral linewidth of the generated radiation.
The coherence times of the ICS source with free $e$-beam and confined $e$-beam are calculated as 27.5\,fs, and 116.2\,fs, respectively.
This means that the $\sim$300 times gain enhancement is obtained at the cost of $\sim$4 times broader bandwidth. 

\section{Radiation energy scaling}

In the previous sections, the drastic increase in the radiated energy caused by the cavity fields was numerically demonstrated.
This section addresses the variation of output energy with respect to the cavity parameters.

\begin{figure}
	\includegraphics[width=3.2in]{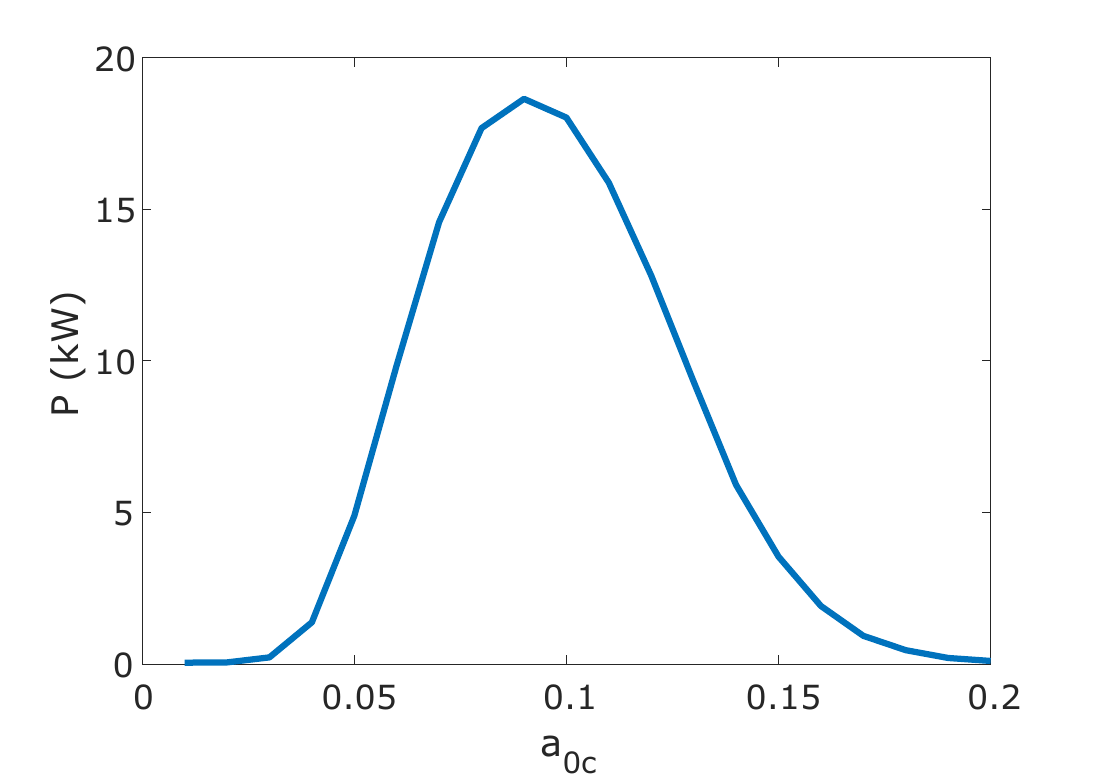}
	\caption{Maximum output power at resonance frequency in the ICS interaction off {color{red}confined} electron beams in terms of the amplitude of the cavity fields, or the so-called beam strength parameter ($a_{0c}$). \label{cavityA0}}
\end{figure}

The first parameter is the beam strength parameter ($a_{0c}$) which determines the ponderomotive force on the particles around the cavity nodes.
This parameter is studied in Fig.\,\ref{cavityA0}, where the output power at resonance frequency is depicted in terms of the beam strength parameter of the cavity.
It is shown that the output power is maximized around a certain ponderomotive force, here $a_{0c} = 0.09$.
In other words, by increasing the ponderomotive force the output power is not necessarily enhanced.
The enhancement takes place merely in the weak ponderomotive force regime.
The reason for this effect is the oscillations caused by cavity fields.
As mentioned previously, the polarization of cavity fields should be selected such that no oscillation happens in the propagation ($z-$) direction, to avoid perturbation in the microbunching process.
This means that the cavity fields wiggle the electrons along the same direction as the ICS laser beam, thereby realizing a pondermotive focusing force.
The wiggling motion due to cavity fields should be negligible compared to those of ICS laser because the main role of the cavity in this scheme is confinement.
As a result, a compromise needs to be reached to maintain strong ponderomotive force and simultaneously weak wiggling motion.
This trade-off is the origin of the results observed in Fig.\,\ref{cavityA0}.

\begin{figure}
	\includegraphics[width=3.2in]{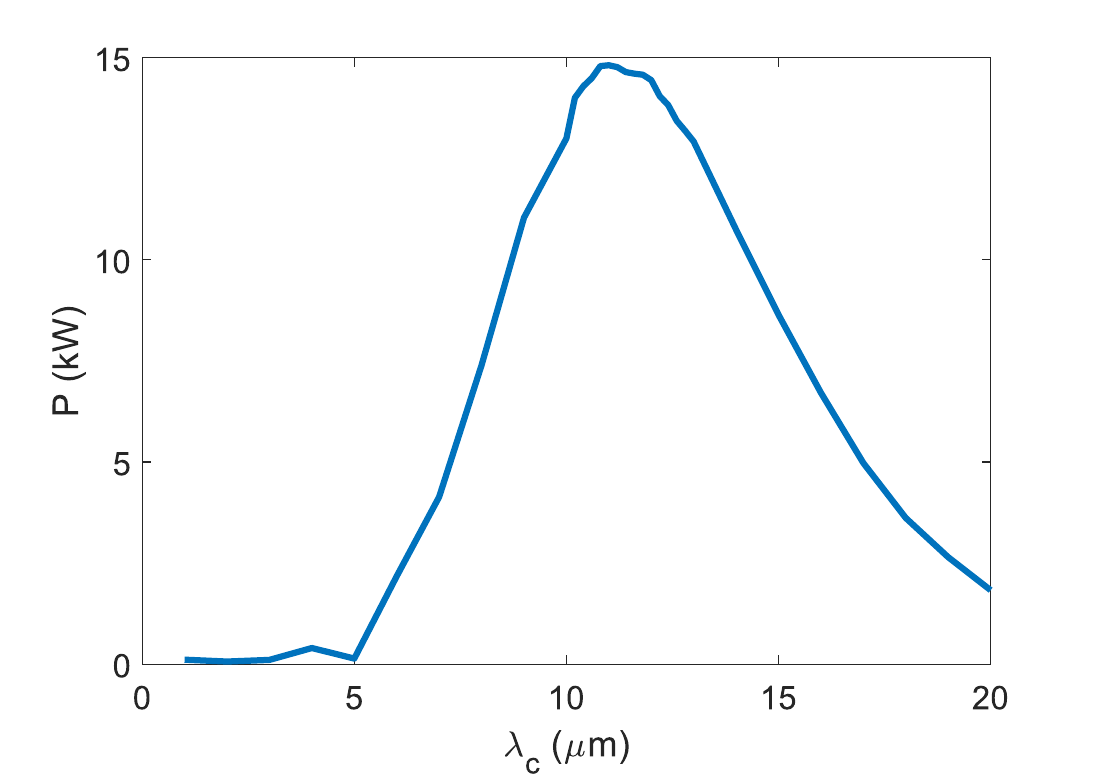}
	\caption{Maximum output power at resonance frequency in the ICS interaction off {color{red}confined} electron beams in terms of the wavelength of the cavity fields ($\lambda_c$). \label{cavityLambda}}
\end{figure}

The next important parameter to be studied is the cavity wavelength, whose influence on the saturation power is examined in Fig.\,\ref{cavityLambda}.
The illustrated simulation results show that the best efficiency is acquired when the cavity wavelength is close to the wavelength of the ICS laser.
The existence of optimal wavelengths for the cavity beams is anticipated because a large wavelength, on one hand, does not impose a considerable confinement effect and a small cavity wavelength, on the other, results in a large number of tiny focal spots that do not ultimately change the bunch transverse profile.
However, the reason why this optimum wavelength occurs at values close to the ICS wavelength is still unknown.
To theoretically prove this, one needs to derive analytical formulations for the radiation of modulated microbeams, which will be presented in a future study.
There exists a small deviation between optimum cavity wavelength and ICS wavelength.
The simulations with $e$-beams of larger transverse sizes showed that this deviation becomes smaller for larger beam sizes.
An anomaly in the radiation at $\lambda_c \approx $ 5\,{\textmu}m is observed in the plot.
This wavelength is equal to the wiggling period of the electrons.
When the two wiggling motions induced by the cavity and ICS laser fields are synchronous, the distribution of the ponderomotive force will be affected and the focal spots will be disturbed.
This leads to the suppression of coherent gain in the radiation generation. 

\section{Conclusion and outlook}

This paper aimed at introducing a new roadmap for realizing compact coherent x-ray sources.
A new scheme based on inverse Compton scattering off low-energy electrons transversely confined inside fields of an optical cavity is presented.
The transverse gradient forces stemming from inhomogeneous cavity fields are capable of confining and additionally focusing the electron beams close to the cavity nodes.
It is shown that the microbeams produced by this effect are more suitable choices for radiation generation compared to a free electron beam.
The possibility of achieving fast coherent gain before bunch expansion due to space-charge forces is theoretically demonstrated.
As a result, the appropriate design of the total interaction leads to compact x-ray sources that enable efficiencies close to the existing x-ray free-electron lasers.

Maximum effort was put into simulating the proposed phenomenon based on first-principle equations in order to capture all involved effects in the interaction between electrons, cavity and counter-propagating beams.
Furthermore, the parameters of electron and optical beams were chosen according to the state-of-the-art $e$-beam and laser technology.
This approach theoretically proves that the proposed scheme may be a viable venue for realizing a ``table-top" coherent x-ray source.
However, the realization of an x-ray source based on the investigated scheme requires a self-consistent start-to-end simulation as well as sensitivity studies.
Particularly, the effects of existing experimental limitations on the source performance need to be scrutinized.
For instance, fluctuations in the phase front of cavity beams, non-uniform (usually Gaussian) current profile in the electron beams, a non-uniform envelope of the ICS laser beam are parameters that influence the efficiency of radiation generation.
In addition, the required synchronization between the electron beam and counter-propagating laser to collide at a certain position with respect to the cavity center needs to be explored.
On the other side, this paper presented the design of a \emph{soft} x-ray source.
The potential of ICS off confined relativistic beams to realize a coherent \emph{hard} x-ray source can be the focus of a future study.
Hopefully, the presented analysis and proposal of one possible solution for a compact coherent x-ray source further motivates this field of research.
Ultimately, these proposals are necessary to achieve a well-performing x-ray source that can be integrated into small labs for easier access and usage by physicists, chemists, and biologists.

\section*{Acknowledgements}

The authors acknowledge the support from the Swiss National Science Foundation (SNSF) for funding the project under the Spark grant CRSK-2 190840.

\section*{Appendix: Standing-wave fields inside an optical cavity}

To consider the effect of particle confinement in cavity fields during an ICS interaction, the same ICS analysis is performed with cavity fields superposed on the fields of the counter-propagating beam.
The fields inside a 2D cavity as shown in Fig.\,\ref{paperConcept} are formulated as a superposition of four Gaussian beams propagating along $+x$, $-x$, $+y$, and $-y$ directions.
These beams produce a standing wave pattern inside the cavity.

In a $(p,s,l)$ Cartesian coordinate system, the fields of a Gaussian beam that is polarized along $p$ direction and propagates along $l$ are formulated as follows:
\begin{widetext}
	\begin{eqnarray}
		E_p &=& \displaystyle E_0 \sqrt{ \frac{ w_{0p} w_{0s} } { w_p(l) w_s(l) } } 
			\exp \left\{ -\frac{p^2}{w_p^2(l)} - \frac{s^2}{w_s^2(l)} \right\} 
			\cos \left( \omega(t - t_0 - \frac{l}{c}) - \frac{k_0p^2}{2R_p(l)} - \frac{k_0s^2}{2R_s(l)} - \frac{\pi}{2} + \frac{\phi_p(l)+\phi_s(l)}{2} \right), \nonumber \\
		E_l &=& \displaystyle E_0 \frac{p w_{0p}}{z_{Rp} w_p(l)} 
		\sqrt{ \frac{ w_{0p} w_{0s} } { w_p(l) w_s(l) } } 
		\exp \left\{ -\frac{p^2}{w_p^2(l)} - \frac{s^2}{w_s^2(l)} \right\}
		\cos \left( \omega(t - t_0 - \frac{l}{c}) - \frac{k_0p^2}{2R_p(l)} - \frac{k_0s^2}{2R_s(l)} + \frac{3\phi_p(l)+\phi_s(l)}{2} \right), \nonumber \\
		B_l &=& \displaystyle \frac{E_0}{c} \frac{s w_{0s}}{z_{Rs} w_s(l)} 
		\sqrt{ \frac{ w_{0p} w_{0s} } { w_p(l) w_s(l) } } 
		\exp \left\{ -\frac{p^2}{w_p^2(l)} - \frac{s^2}{w_s^2(l)} \right\}
		\cos \left( \omega(t - t_0 - \frac{l}{c}) - \frac{k_0p^2}{2R_p(l)} - \frac{k_0s^2}{2R_s(l)} + \frac{\phi_p(l)+3\phi_s(l)}{2} \right), \nonumber \\
		B_s &=& \frac{E_p}{c}, \qquad E_s = B_p = 0 \nonumber
	\end{eqnarray}
where
\begin{equation}
	\phi_\zeta(l) = \tan^{-1}(l/z_{R\zeta}), \qquad
	w_\zeta(l) = w_{0\zeta}\sqrt{1+(l/z_{R\zeta})^2}, \qquad
	R_\zeta(l)=l(1+(z_{R\zeta}/l)^2), \nonumber
\end{equation}
\end{widetext}
with $\zeta \in \{s,p\}$.
$w_{0p}$ and $w_{0s}$ represent the beam radia parallel and normal to the polarization vector, respectively.
$z_{Rp} = k_0 w_{0p}^2 / 2 $ and $z_{Rs} = k_0 w_{0s}^2 / 2 $ are the Rayleigh range values parallel and normal to the polarization vector, respectively.
$t_0$, $\omega$ and $k_0=\omega/c$ are the time delay, angular frequency and wave number of the beam, respectively.

\end{document}